\documentclass[aps,prl,twocolumn,groupedaddress,notitlepage,showpacs,floatfix,superscriptaddress]{revtex4-1}
\pdfoutput=1
\usepackage{graphicx,graphics,epsfig,subfigure,times,bm,bbm,amssymb,amsmath,amsthm,mathrsfs,MnSymbol}
\usepackage{gensymb}
\usepackage{amsfonts}
\usepackage[matrix,frame,arrow]{xypic}
\usepackage[pdfstartview=FitH]{hyperref}
\hypersetup{
    colorlinks=true,       
    linkcolor=red,          
   citecolor=magenta,        
    filecolor=magenta,      
    urlcolor=cyan,           
    runcolor=cyan
}
\usepackage[pdftex]{color}
\usepackage{hypernat}
\usepackage{braket}
\usepackage{enumerate}
\usepackage[normalem]{ulem}
\usepackage[usenames,dvipsnames]{xcolor}
\usepackage{multirow}
\usepackage{mathtools}

\definecolor{orange}{rgb}{1,0.5,0}

\newcommand{\ignore}[1]{}
\usepackage{geometry}

\ignore{
\documentclass[eprintnumbers,amsmath,amssymb,onecolumn,a4paper,caption ]{article}
\usepackage{amsfonts}
\usepackage{amssymb}
\usepackage{mathrsfs}
\usepackage{mathbbold}
\usepackage{bbm}
\usepackage{mathrsfs}
\usepackage{dcolumn}
\usepackage{bm}
\usepackage{times,epsfig,amssymb,amsmath}
\usepackage{float}
\usepackage{subfigure}
\usepackage{geometry}\geometry{left=2.5cm,right=2.5cm,top=3cm,bottom=3cm}

\usepackage{color}

}

\begin{document}

\title{Disorder-protected topological entropy after a quantum quench }

\author{Yu~Zeng}
\affiliation{Beijing National Laboratory for Condensed Matter Physics, Institute of Physics, Chinese Academy of Sciences, Beijing 100190, China}
\affiliation{School of Physical Sciences, University of Chinese Academy of Sciences, Beijing 100190, China}
\author{Alioscia~Hamma}
\thanks{Alioscia.Hamma@umb.edu}
\affiliation{Department of Physics, University of Massachusetts Boston, 100 Morrissey Blvd, Boston MA 02125}
\author{Heng~Fan}
\affiliation{Beijing National Laboratory for Condensed Matter Physics, Institute of Physics, Chinese Academy of Sciences, Beijing 100190, China}
\affiliation{School of Physical Sciences, University of Chinese Academy of Sciences, Beijing 100190, China}
\affiliation{Collaborative Innovation Center of Quantum Matter, Beijing 100190, China}
\begin{abstract}
Topological phases of matter are considered the bedrock of novel quantum materials as well as ideal candidates for quantum computers that possess robustness at the physical level. The robustness of the topological phase at finite temperature or away from equilibrium is therefore a very desirable feature. Disorder can improve the lifetime of the encoded topological qubits. Here we tackle the problem of the survival of the topological phase as detected by topological entropy, after a sudden quantum quench. We introduce a method to study analytically the time evolution of the system after a quantum quench and  show that disorder in the couplings of the Hamiltonian of the toric code and the resulting Anderson localization can make the topological entropy resilient.

\end{abstract}


\maketitle




{\em Introduction.---}
Novel quantum phases in many-body systems that feature topological order are of extreme importance in both condensed matter physics \cite{wenbook} and in quantum information \cite{nayak:2008}. They possess gapped energy spectrum and robust ground-state degeneracy, which is supposed to be a promising candidate of the self-correcting quantum memory \cite{dennis}.
These novel quantum phases can not be described by the Landau paradigm of symmetry breaking and are not characterized by local order parameters. Instead, they are characterized by a long-range pattern of entanglement dubbed topological entropy (TE) \cite{hiz1, hamma:2005b, kitaevpreskill,levin:2006} that serves as nonlocal order parameter \cite{hammahaas}.

In order to exploit topological order for realistic applications such as the robust quantum memory, the system needs to be robust not only in the ground state degeneracy but must also feature robustness at both the dynamical level and at finite temperature \cite{chesitherm, iblisdir, chamon3d}. Topologically ordered systems in two and three dimensions based on local Hamiltonians with commuting operators are not stable both at finite temperature \cite{rmp2016,Hastings2011,finiteT,chamon3d} or when cast away from equilibrium \cite{kay2009,Yu2016}. On the other hand,   both the topological phase and its self correcting quantum memory are  robust in four or greater spatial dimensions, which, unfortunately, is not realistic for implementation \cite{dennis, alicki, rmp2016,hammamazac}. The depletion of both topological entropy and topological quantum memory is due to the diffusion of defects that ultimately destroy both features, as they are intimately connected, although not exactly the same thing\cite{chamon3d, hammamazac, nussinov}. Several schemes have been proposed to overcome these shortcomings, from the introduction of long-range interactions between the excitations \cite{ toricboson, chesi-mem, Bardyn2016}, to models that feature membrane condensation \cite{membrane} together with the absence of string-like excitations \cite{haah}, and the introduction of localization through disordered couplings \cite{disorderTCM1, disorderTCM2, bravyi_majorana},  the latter showing that disorder can increase the lifetime of quantum memory,  also see \cite{rmp2016} for extended references.

In this paper, we study how disorder can improve the resilience of topological order after a quantum quench. To this end, we   study the time evolution of TE in the toric code with randomized couplings. While for the clean system the TE will self-thermalize after the quantum quench \cite{Yu2016},  we show that we can obtain a stable TE with a error that can be made arbitrarily small by a disorder increasing { nearly} with the square root of the system size.

\begin{figure}
\centering
\includegraphics[width=0.4\textwidth]{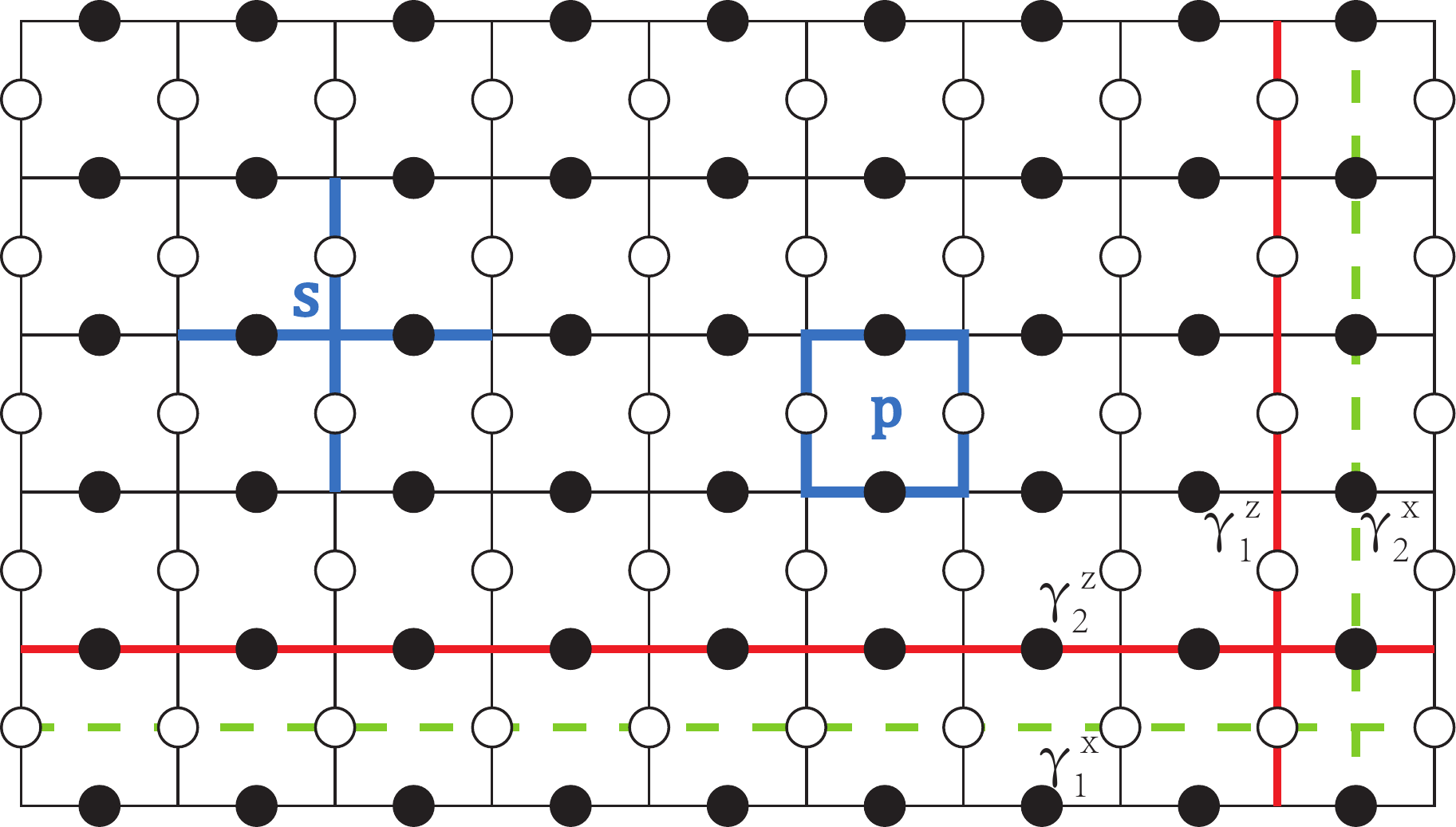}\\
\caption{(Color online) Illustration of the square lattice with physical spins living on the bonds in odd rows (black dots) and even rows (white dots). The Examples of star (s), plaquete (p), and the non-contractible path $\gamma^\alpha_a$ ($\alpha=x,z$ and $a=1,2$) are shown.}\label{lattice}
\end{figure}

{\em Quantum quench and time evolution of TE.---}
 We consider the two-dimensional toric code model (TCM) introduced by Kitaev \cite{kitaev:2003} defined on a periodic $M\times N$ rectangular lattice, with spins $1/2$ on the bonds. The TCM Hamiltonian is given by
 \begin{eqnarray}\label{HTC}
 H_{TC}=-\sum_{s}A_{s}-\sum_{p}B_{p}
 \end{eqnarray}
where the star operators $A_{s}\equiv\prod_{i\in s}\sigma_{i}^{x}$ and the plaquette operators $B_{p}\equiv\prod_{i\in p}\sigma_{i}^{z}$ are stabilizer operators which belong to stars($s$) and plaquettes ($p$) on the lattice containing four spins each (see Fig. \ref{lattice}). The Hamiltonian is exactly solvable since any two stabilizer operators commute. So the ground space of the Hamiltonian is made of  simultaneous eigenstates of all  stabilizer operators with eigenvalue $+1$. Considering the global constrains $\prod_{s}A_{s}=\prod_{p}B_{p}=\mathbbm{1}$, one can see  that the ground-state manifold $\mathscr{L}$ is fourfold degenerate. The logical operators encoding the topological qubits are given by ($W^{x}_{1},W^{z}_{1}$) and ($W^{x}_{2},W^{z}_{2}$) where $W^\alpha_a$ is defined as $W^\alpha_a=\prod_{j\in\gamma^\alpha_a}\sigma^\alpha_a$ ($\alpha=x,z$ and $a=1,2$) with each $\gamma^\alpha_a$  a non-contractible string winding around the torus (see Fig. \ref{lattice}). By defining the reference state $|0\rangle=1/\sqrt{2^{NM-1}}\prod_s(1+A_s)|\Uparrow\rangle$, with $|\Uparrow\rangle$ being the all spin-up state in the $\sigma^z$ basis, a generic state in the ground state manifold  can be expressed as  $|\xi\rangle=\sum_{i,j=0}^{1}a_{ij}(W_1^{x})^{i}(W_2^{x})^{j}|0\rangle$ with $\sum_{i,j=0}^{1}a_{ij}^2=1$.

As the TCM system is gapped,  the bipartite entanglement in the ground state satisfies the area law for the entanglement \cite{arealaw, amico}. However, there is a $O(1)$ correction of topological origin \cite{hiz1} that can be extracted by a clever linear combination of different entropies corresponding to different subsystems \cite{kitaevpreskill,levin:2006}. This topological correction is a stable feature of the topological quantum phase \cite{hammahaas, santra2014}. Remarkably, the same linear combination obtained by the R\'enyi entropies also serves as order parameter for the topological phase \cite{flammia:2009, halasz} and it is easier to compute and experimentally accessible in principle \cite{Abanin2012,Zoller2012}, especially for the case of 2-R\'enyi entropy \cite{nature15750}: $S^{R}_{2}=-\log_{2}\mbox{Tr}[\rho^2_{R}]$, where $P=\mbox{Tr}[\rho^2_{R}]$ is the purity of the state reduced to the subsystem $R$, namely $\rho_R$.

\begin{figure}
\centering
\includegraphics[width=0.35\textwidth]{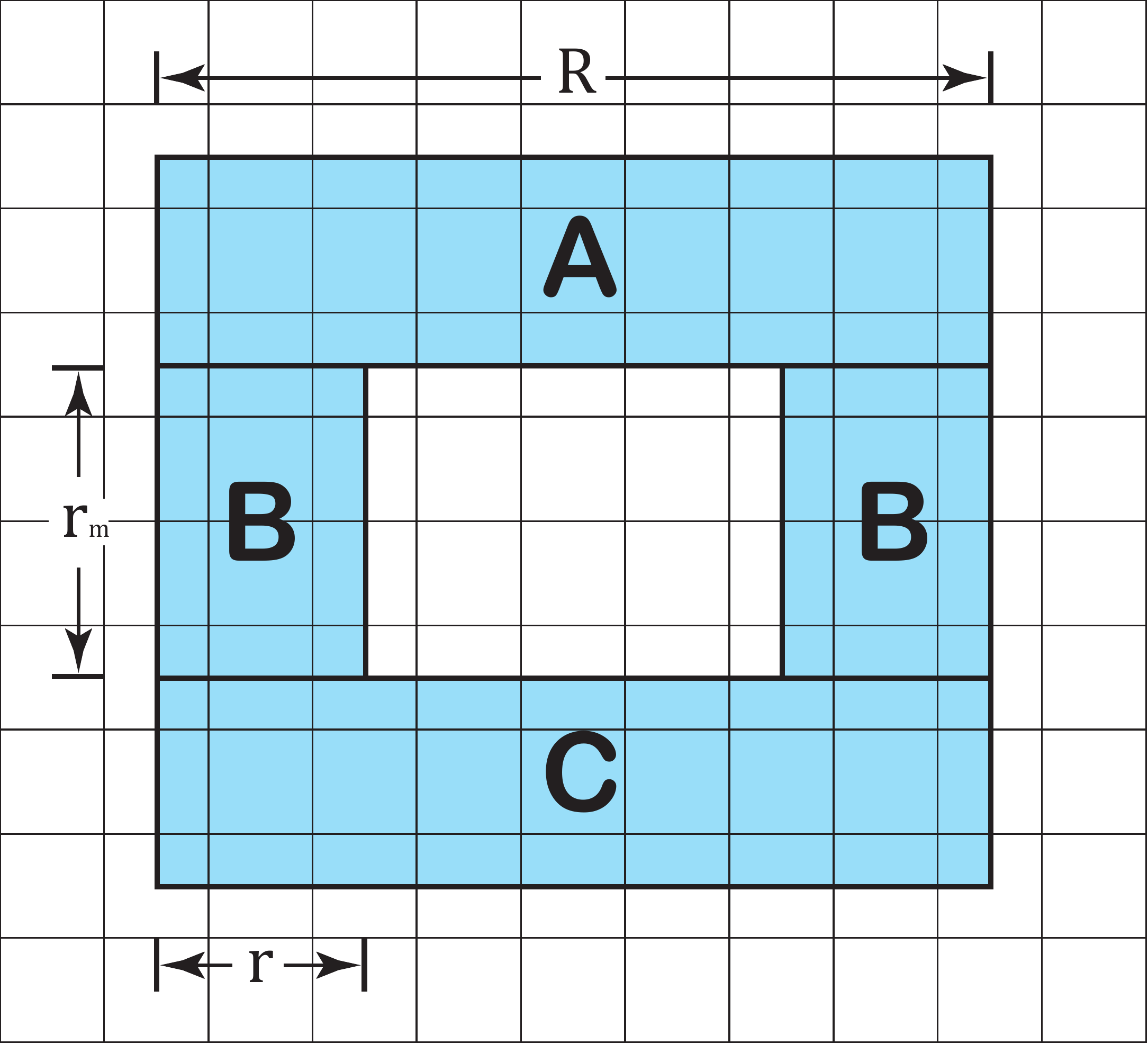}\\
\caption{(Color online) Illustration of the subsystems $A$,$ B$ and $C$ with extension $R$ and thickness $r$. { The number of lattice points in horizontal direction} is $N=200$ in the evaluation of the analytic result. The distance between $A$ and $C$ is $\text{r}_m$ and can scale with { the number of lattice points in the vertical direction, $M$,} while $r=2,R=8$ .}\label{4regions}
\end{figure}

One can directly calculate the von Neumann entropy\cite{hamma:2005b} and the 2-R\'enyi entropy \cite{flammia:2009, santra2014}  for a simply connected region $R$  of an arbitrary state in ground-state manifold of TCM to obtain $S^{R}_{von}=S^{R}_{2}=L_R-1$. Here $L_R$ is the boundary length of the region $R$  while the sub-leading constant is topological entropy characterizing the topological phase. Following  Ref.\cite{levin:2006, flammia:2009}, the topological R\'enyi entropy is defined as $S^{T}_{\alpha}=S^{AB}_{\alpha}+S^{BC}_{\alpha}-S^{B}_{\alpha}-S^{ABC}_{\alpha}$, where the subsystem A, B and C are illustrated in Fig.\ref{4regions}. In the ground state of TCM, we have $S^T_2=2$.

In order { to} explore the non-equilibrium time evolution of $S^{T}_2$ at zero temperature, we study the scenario of the sudden quantum quench. We prepare the initial quantum memory as one ground state of TCM at t=0: $|\Psi(0)\rangle=1/\sqrt{2}(1+W^x_1)|0\rangle$ \cite{note1}, then we suddenly change the Hamiltonian to
\begin{eqnarray}\label{Hsigma}
\!\!\!\!\!H(J,\lambda)\!\!=\!-\!\sum_{s}J_{s}A_{s}\!-\!\sum_{p}J_{p}B_{p}\!-\!\!\!\sum_{\substack{i\in \text{odd}\\\text{rows}}}\!\!\lambda^{z}_{i}\sigma^{z}_{i}\!-\!\!\!\sum_{\substack{j\in \text{even}\\\text{rows}}}\!\!\lambda^{x}_{j}\sigma^{x}_{j}.
\end{eqnarray}
$J_s$ and $J_p$ are the stabilizer strengths depending on their space positions, which is random in general. The external fields are arranged in a special fashion (see Fig.~\ref{lattice}). Superimposing together the original lattice and the dual lattice, we place the field in the $z$ direction with magnitude of $\lambda^{z}_{i}$ on the odd rows (black dots) and the field in the $x$ direction with $\lambda^x_{j}$ on the even rows (white dots). The wave function will undergo a unitary time evolution $|\Psi(t)\rangle=e^{-itH(J,\lambda)}|\Psi(0)\rangle$. Note that $|\Psi(t)\rangle$ is always in the sector of $W^x_1=1$ and $W^z_2=1$ \cite{note1}. The time-dependent topological 2-R\'enyi entropy is given by
\begin{eqnarray}\label{ST2}
 S^T_2(t,J,\lambda)=log_2\left(\frac{P^{ABC}(t,J,\lambda)P^{B}(t,J,\lambda)}{P^{AB}(t,J,\lambda)P^{BC}(t,J,\lambda)}\right)
\end{eqnarray}
and therefore the behavior of $S^T_2(t)$ is determined by the purity for each subsystem, namely, $P^{R}(t)=\mbox{Tr}_R \left[ \mbox{Tr}_{\bar{R}}\left( e^{-iHt }  |\Psi(0)\rangle\langle\Psi(0)| e^{iHt } \right)\right]^2$.

The Hamiltonian Eq.(\ref{Hsigma}) can be divided into two mutual commutative parts ${H=H_1+H_2}$, where $H_1=-\sum_{s}J_{s}A_{s}-\sum_{i\in \text{odd}~\text{rows}}\lambda^{z}_{i}\sigma^{z}_{i}$ and $H_2=-\sum_{p}J_{p}B_{p}-\sum_{j\in \text{even}~\text{rows}}\lambda^{x}_{j}\sigma^{x}_{j}$. We can map the stabilizer operators to the effective spins living on the lattice and dual-lattice sites, which means $A_s\mapsto\tau^z_s$ and $B_p\mapsto\tau^z_p$. In this spin `$\tau$-picture', the external fields $\sigma^z_i$ and $\sigma^x_j$ flip their two neighbour effective spins, thus $\sigma^z_i\mapsto\tau^x_s\tau^x_{s^{\prime}}$ and $\sigma^x_j\mapsto\tau^x_p\tau^x_{p^{\prime}}$, where i labels the bond between two neighboring sites $\langle s,s^{\prime}\rangle$ on the lattice while j labels the bond between $\langle p,p^{\prime}\rangle$ on the dual-lattice. The corresponding Hamiltonian of Eq.(\ref{Hsigma}) in the `$\tau$-picture' is the sum of total $2M$ independent Ising chains with periodic boundary conditions in either odd or even rows:
\begin{eqnarray}\label{Htau}
\tilde{H}(J,\lambda)&=&\sum_{\substack{s\in \text{odd}\\\text{rows}}}\left(-J_s\tau^z_s-\lambda^z_{\langle s,s^{\prime}\rangle}\tau^x_s\tau^x_{s^\prime}\right)\nonumber\\
&+&\sum_{\substack{s\in \text{even}\\\text{rows}}}\left(-J_p\tau^z_p-\lambda^z_{\langle p,p^{\prime}\rangle}\tau^x_s\tau^x_{s^\prime}\right),
\end{eqnarray}
{ In presence of random couplings $J_s, J_p$, this model features Anderson localization for the $\tau$ degrees of freedom, which correspond to the anyonic excitations of the toric code} {\cite{Anderson1958, kitaev:2003}}.
Mapping into the `$\tau$-picture', the initial state $|\Psi(0)\rangle\mapsto|\tilde{\Psi}(0)\rangle$ turns out to be the all spin-up state and the time-evolution state $|\Psi(t)\rangle\mapsto|\tilde{\Psi}(t)\rangle$ results in the tensor-product state of each rows. { 

Although  in the $\tau$-picture the state can be expressed as the tensor product of the {$2M$} rows, that does not imply that the system is not entangled in two dimensions in the original spin degrees of freedom, namely the  $\sigma$-picture. We employ the $\tau$-picture as a tool to compute correlation functions that enter in the computation of $S^T_2$, as it was shown in \cite{Yu2016}. In the $\tau$-picture, the formula for the purity of any subsystem $R$ reads as
\begin{eqnarray}\label{purity}
\!P(t)\!=&C_{P}&\!\!\sum_{\partial\tilde{g}\in \partial G^\prime_{R}} \sum_{\substack{\tilde{g}\in G^\prime_R\\ \tilde{z}\in Z_R}}\!\!|\langle\tilde{g}\partial\tilde{g}\tilde{z}\rangle_{\Phi_1\!(\!t\!)}|^2\!\!\sum_{\substack{\tilde{h}\in H^\prime_R\\ \tilde{x} \in X^{\prime}_{R}}}\!\!|\langle\tilde{x}\partial\tilde{x}(\partial\tilde{g})\tilde{h}\rangle_{\Phi_2\!(\!t\!)}|^2\nonumber\\
&\times&\sum_{\partial\bar{g}\in \partial G^\prime_{\bar{R}}}(-1)^{\partial\bar{g}\partial\bar{x}(\partial\bar{g})\cap\tilde{z}\tilde{h}}
\end{eqnarray}
In the above formula, $\Phi_1(t)$ and $\Phi_2(t)$ describe the time evolution of the system in odd rows and even rows respectively. The operators $\tilde{g}, \tilde{z}, \tilde{h}, \tilde{x}, \partial\tilde{g}\partial\tilde{x}(\partial\tilde{g})$ and $\partial\bar{g}\partial\bar{x}(\partial\bar{g})$ represent string operators in `$\tau$-picture' operating with the Pauli algebra in either subsystem $A$ or its complement. The definition of these operators is detailed in the supplemental material. The phase factor takes into account whether the two operators commute or anti-commute. The constant $C_{P}$ is irrelevant when we compute the $S^T_2$ by Substituting Eq.(\ref{purity}) into Eq.(\ref{ST2}). As one can see, the calculation requires all the knowledge of the many-spin correlation functions in the summation. In the supplemental material, one can find the details for the  calculation of all these correlation functions, even for the disordered system. Through mapping to free fermions, the calculation of  correlation functions is mapped onto the evaluation of a Pfaffian,  {which can be reduced to a determinant whose maximal dimension is $2(R+1)$} \cite{barouch:1971}. Only the evaluation of the Pfaffian has to be performed by a computer. The complexity of the problem only resides on the fact that Eq.(\ref{purity}) contains a number of correlation functions that is exponential in  { $(R+r)$}, but each of these correlation functions can be evaluated very efficiently.

{\em Main result.---}
After this effort, one can obtain the value of the purity Eq.(\ref{purity}) and the topological 2-R\'enyi entropy Eq.(\ref{ST2}). For the clean case, that is $J_s=J_p=1$ and $\lambda_s=\lambda_p\neq0$, the system is integrable, and  $S^T_2(t)$ vanishes in long time evolution after a quantum quench therefore reaching the value of thermal equilibrium \cite{Yu2016}. At long times, the encoded logical qubit will be lost. Moreover,  the  Loschmidt echo $|\langle\Psi (t)|\Psi (0)\rangle |^2$ decays \cite{Yu2016} together with the effective magnetization in TCM \cite{kay2011}.

We have mentioned that previous results \cite{rmp2016} indicated an improvement of the lifetime for the topological encoded qubit. Now we set out to show our main result, that the dynamical localization of the anyons due to the disorder in the couplings makes the TE resilient. As we shall see, it is very important to see how disorder must scale with the system size and with the amount of protection desired.

The coupling strengths  $J_s=1+\delta J_s$ and $J_p=1+\delta J_p$ are randomized as $\delta J_s$ and $\delta J_p$ are  uniformly distributed   in $[-W_J, W_J]$. The external-field strengths are set to be fixed $\lambda_s=\lambda_p=0.6$.  We set the system size as $N=200$, and the subsystem with extension $R=8$ and thickness $r=2$. On the other hand, $r_m$ is made to scale and $M$ can be any number proportional to $r_m$.  Given a disorder strength $W_J$, the resulting TE as a function of time $S^T_2(t)$ were averaged over 100 realizations of the disorder. A simplification (without loss of generality) is used here, that each row has the same arrangement of stabilizer strengths in every single realization, so to reduce the total number of correlation functions to compute.

The results are displayed in Fig. \ref{disorderquench}, where we show the time evolution of $S^T_2(2)$ for $r_m=9$, and different disorder strengths $W_J$ from $0$ to $50$ in interval of $5$.  We can see that after a very short time the average  $S^T_2(t) $ tends to equilibrate with small fluctuations. As the disorder strength $W_J$ increases, the equilibrium value $S^T_2(t)$ is closer to the initial value. The main message of the result is that Anderson localization induced by the disordered stabilizer strength makes the topological order resilient after a quantum quench. 

\begin{figure}
\centering
\includegraphics[width=0.48\textwidth]{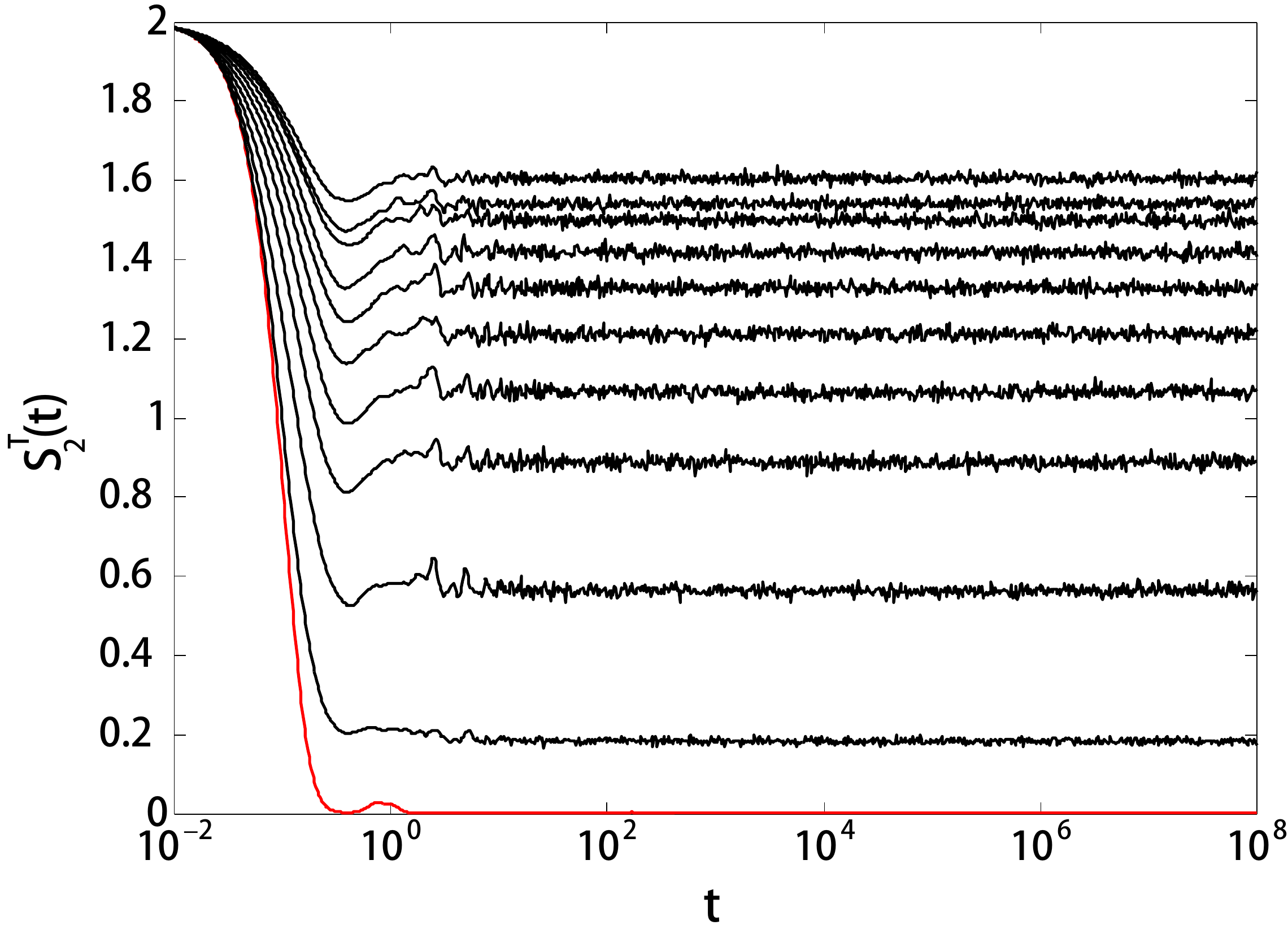}\\
\caption{(Color online)  Illustration of $S^T_2(t)$ after a quantum quench with different disorder strengths. The distance between subsystem A and C is $r_m=9$. The 11 distinct curves from bottom to to correspond to $W_J$ equal to 0 (red curve), 5, 10, $\cdots$, 45, 50. }\label{disorderquench}
\end{figure}

One here needs to be careful with the scaling of these quantities with the system size. It turns out that the dependence is on the size of the
`hole' $r_m$ in the way the subsystem is partitioned. We define the quantity
\begin{eqnarray}
\Delta S^T_2\equiv S^T_2(0)-\lim_{t\rightarrow\infty}S^T_2(t).
\end{eqnarray}
indicating the drop of TE after infinite time. As $r_m$ increases, this drop also increases, as it is shown in
Fig. \ref{DeltaS2}. 
However, we can see that as the disorder is increased, one can make this drop arbitrarily small. To determine the behavior of $\Delta S^T_2$ with large $r_m$ and disorder strength $W_J$, we perform a scaling collapse with a scaling function taking the form
\begin{eqnarray}
\Delta S^T_2(r_m,W_J)=g(r_m)f(W_J r^{-b}_m),
\end{eqnarray}
where $b$ is the scaling parameters, and $f$ is an undetermined function. We choose { $g(r_m)=\frac{1}{2}\Delta S^T_2(r_m,W_J=0)$}. Numerical fitting shows that $g(r_m)$ satisfies a exponential function: { $g(r_m)=\frac{1}{2}(q_0e^{-p_0r_m}+a_0)$}, where $q_0=-2.73$, $ p_0=0.91$, and  $a_0=2$. The inset to Fig. \ref{disorderquench} shows the collapsed data and the obtained scaling parameter, where $x_{r_m}=W_J r^{-b}_m$ and  $y_{r_m}=\Delta S^T_2/g(r_m)$. The collapsed data shows that in order to keep TE unchanged when $r_m$ increases, one needs to increase disorder strength grows as $W_J\sim r^{b}_m$, where $b=0.48$.
\begin{figure}
\centering
\includegraphics[width=0.48\textwidth]{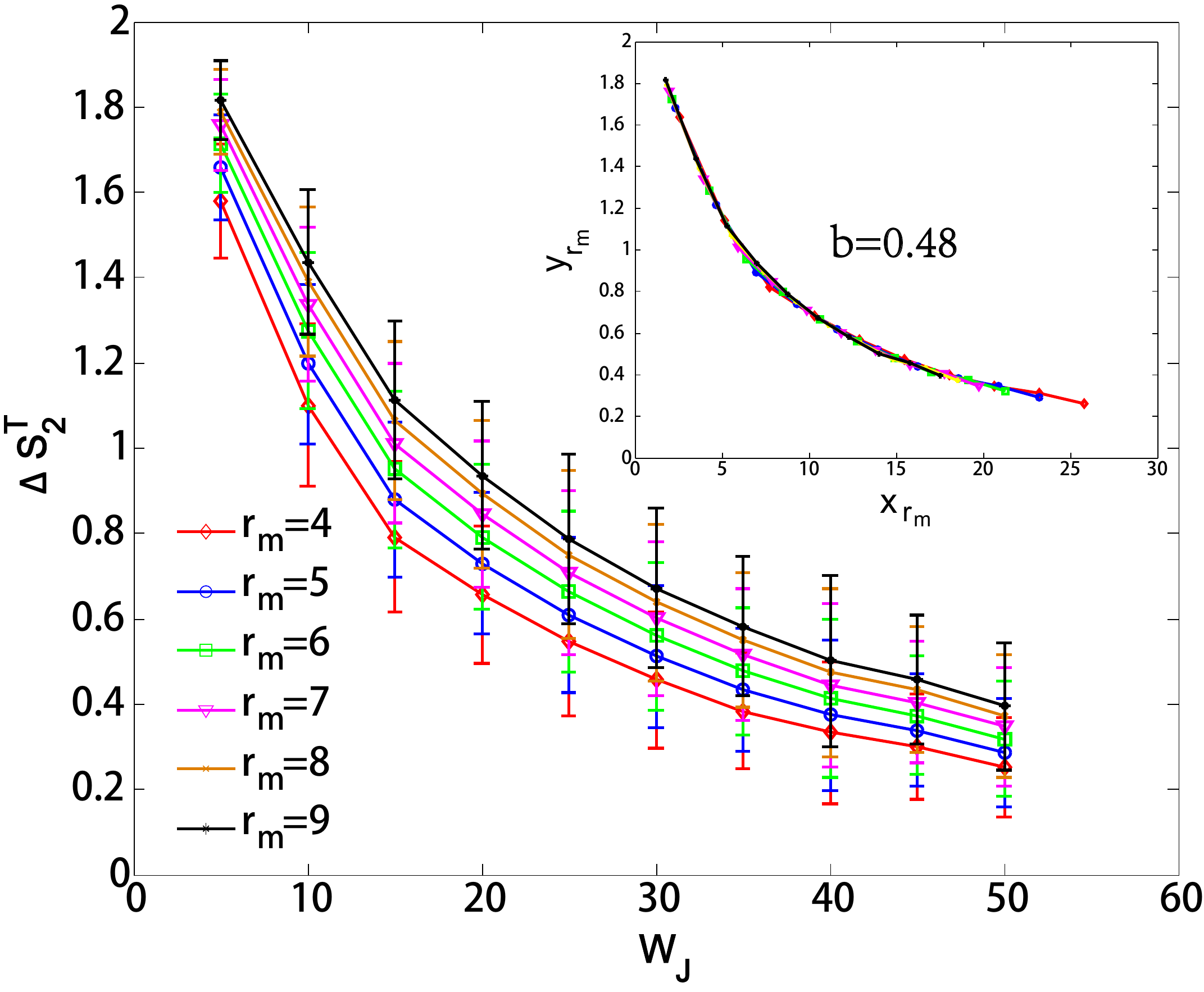}\\
\caption{(Color online) $\Delta S^T_2$ as a function of disorder strength $W_J$ with distinct $r_m$. The inset gives the scaling collapse of the data.}\label{DeltaS2}
\end{figure}
{For $x_{r_m}\gg 1$}, the fitted function $f$ takes the form { $f(x_{r_m})=16.9/(x_{r_m}^{1.23}+7.23)$}, which shows that one can obtain an arbitrarily good protection of the TE by setting $\Delta S^T_2(r_m,W_J)<\epsilon$ and then scaling the disorder strength with { $r_m^bh(\epsilon)$, where $h(\epsilon)=(16.9/\epsilon-7.23)^{0.81}$ from the fitted funtion}.

{\em Conclusions and Outlook.---}
In this paper, we investigated the resilience of the topological phase after a quantum quench by means of dynamical localization induced by disorder in the couplings of the Hamiltonian. We use the topological entropy as a order parameter for the phase. In the toric code with random couplings and a special arrangement of the external fields, the system can be cast in the form of free fermions, and disorder will induce
Anderson localization for the anyonic excitations of the system. We have shown that the phase can be protected arbitrarily well at arbitrarily long times by scaling the disorder strength with the square root of the system size.

 If we  let the  external fields in Eq.(\ref{Hsigma}) take a more general form, an   interacting terms for the fermions appears. Thus strong disorder should   give rise to many-body localization (MBL) \cite{huse1,gogolin, Bloch15,Choi16,Monroe16}. MBL features a very slow dynamic and absence of thermalization at reasonable times \cite{abanin2, abaninlog, junyang} and preserves initial information to some extent. With low disorder, the system is supposed to thermalize. However, if the interaction strength corresponding to the external field is small enough, the localization still trumps the propagation \cite{Basko2005}. So it is possible  that the protection of topological phase will  hold for small enough general external field. The competition of the disorder strength in the  stabilizer couplings and interaction strength from external field leads to the MBL phase transition. The crossover study on topological order and MBL phase transition will be the scope  of future research.

\begin{acknowledgments}
This work was supported by MOST of China (Grants No. 2016YFA0302104 and
2016YFA0300600), national natural science foundation of China NSFC (Grant No. 91536108)
and Chinese Academy of Sciences (Grants No.
XDB01010000 and XDB21030300) (H.~F.).
\end{acknowledgments}

%

\section{Supplemental material}
\subsection{Definition of the operators in the purity formula}
In this section we provide the definition of the groups of string operators in the purity formula Eq.(\ref{purity}) in the main text, namely $\partial G^{\prime}_R$, $G^\prime_R$, $Z_R$, $H^\prime_R$, $X^\prime_R$ and $\partial G^\prime_{\bar R}$. Considering the mapping from the physical spins, namely `$\sigma$-picture', to the `$\tau$-picture' is isotropic in the Hilbert space we concern, we will not distinguish the notations in the two pictures. We first define the operators in the `$\sigma$-picture', then map them to the `$\tau$-picture' for  further calculation \cite{Yu2016}.

Given a subsystem $R$, $X^\prime_R$ is the group generated by the $\sigma^x$ operators  on the lattice bonds belonging to $R$  (on the even rows). $G^\prime_R$ is formed by the products of star operators, such that every element of the group acts only on the subsystem $R$. For a compact lattice, One can also define $X^\prime_{\bar{R}}$ and $G^\prime_{\bar{R}}$ for the complement subsystem of $R$ in the same fashion. $H^\prime_R$ is formed by all the products of plaquette operators which commute with any element in group $X^\prime_{\bar{R}}$. $Z_R$ is generated by the $\sigma^z$ operators on the odd rows which commute with any element in $G^\prime_{\bar{R}}$. As an example, we show the generators of $X^\prime_R$ and $Z_R$ in Fig. \ref{SM_XZfig} for the subsystem being $AB$, where each white (black) dot represents a $\sigma^x$ ($\sigma^z$) operator. So far, as the element in each group, $\tilde{z}\in Z_R$, $\tilde{x}\in X^\prime_R$, $\tilde{g}\in G^\prime_R$ and $\tilde{h}\in H^\prime_R$ are well defined.

The definitions of $\partial\tilde{g}\partial\tilde{x}(\partial\tilde{g})$ and $\partial\bar{g}\partial\bar{x}(\partial\bar{g})$ are more complicated, so let us do it step by step. First, we denote by $\partial\tilde{g}$  a product of star operators  that does not belong to either $G^\prime_{R}$ or $G^\prime_{\bar{R}}$. This means that $\partial\tilde{g}$  acts on both $R$ and $\bar{R}$. Second, $\partial\tilde{x}$ is defined as an element of $X^\prime_{\bar{R}}$. Third, $\partial\tilde{x}$ is a function of $\partial\tilde{g}$, such that $\partial\tilde{g}\partial\tilde{x}(\partial\tilde{g})$ acts only on the subsystem $R$. And finally, $\partial\tilde{g}\partial\tilde{x}(\partial\tilde{g})$ can not belong to the group $G^\prime_R\times X^\prime_{R}$. The set of the $\partial\tilde{g}\partial\tilde{x}(\partial\tilde{g})$'s satisfying all the above four conditions forms a group. Because $\partial\tilde{x}$ is a function of $\partial\tilde{g}$, This group is mapped to $\partial G^{\prime}_R$ by $\partial\tilde{g}\partial\tilde{x}(\partial\tilde{g})\mapsto\partial\tilde{g}$. We can get $\partial\bar{g}\partial\bar{x}(\partial\bar{g})$ and $\partial G^{\prime}_{\bar{R}}$ by changing $R$ to $\bar{R}$.
\begin{figure}
\centering
\includegraphics[width=0.40\textwidth]{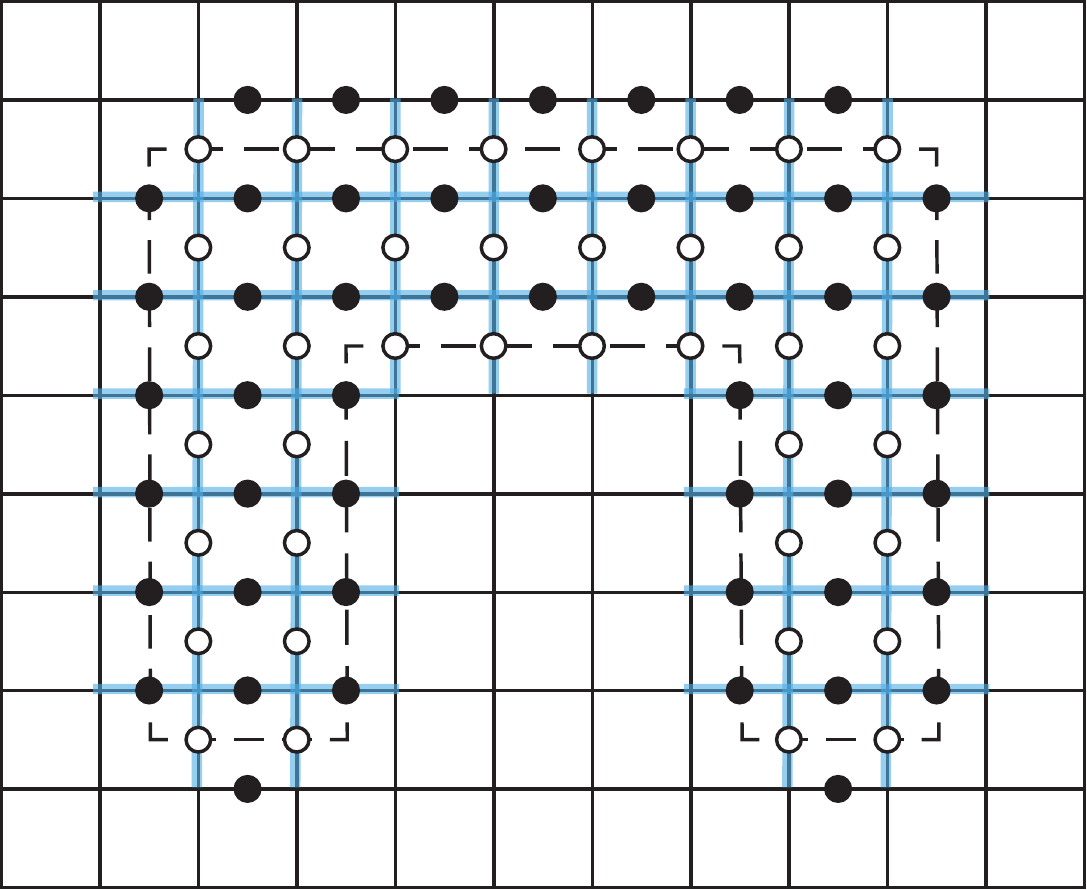}\\
\caption{(Color online) $\sigma^x$($\sigma^z$) operators on the white (black) dots generate the group $X^\prime_R$ ($Z_R$) for subsystem $R=AB$ showed by the blue segments.}\label{SM_XZfig}
\end{figure}

\subsection{ Calculation of the correlation functions}
In this section, we show the details of the calculation of the correlation functions necessary for the evaluation of the purity Eq.(\ref{purity}).
The time evolved state after the quantum quench is expressed as a tensor-product over all the rows  in the $\tau$-picture. Now we show how to find the state for every row,  by mapping each Ising chain into free fermions through Jordan-Wigner transformation.
We consider an arbitrary row and label the `$\tau$' spins with $l$ from $1$ to $N$. The corresponding  Ising Hamiltonian with random couplings \cite{randomising} is
\begin{eqnarray}
\tilde{\mathscr{H}}=-\sum_{l=1}^N\left(J_{l}\tau^{z}_{l}+\lambda_{l}\tau^{x}_{l}\tau^{x}_{l+1}\right).
\end{eqnarray}
The Jordan-Wigner transformation maps the spin operators into the fermion operators by $\tau^z_l=1-2c^{\dagger}_lc_l$ and $\tau_l^{+}=(\tau_l^{-})^\dagger=\prod_{j=1}
^{l-1} (1-2c^{\dagger}_j c_j)c_l$. We can rewrite the above Hamiltonian as
\begin{eqnarray}\label{fh}
\!\!\tilde{\mathscr{H}}=&&-\sum_{l=1}^N\!J_l(c_l^\dagger\!+\!c_l)(c_l^\dagger\!-\!c_l)\!-\!\!\sum_{l=1}^{N-1}\!\!\lambda_l(c_l^\dagger\!-\!c_l)(c_{l+1}^\dagger\!+\!c_{l+1}\!)\nonumber\\
&&+\lambda_{N}\text{exp}(i\pi\sum_{l=1}^N c_l^\dagger c_l)(c_N^\dagger-c_N)(c_1^\dagger+c_1).
\end{eqnarray}
 Note that  in the fermion representation the all spin-up state is mapped to the vacuum state  denoted as $|\tilde{\psi}(0)\rangle$, which correspond to the initial state. The state evolves as $|\tilde{\psi}(t)\rangle=e^{-it\tilde{\mathscr{H}}}|\tilde{\psi}(0)\rangle$, and the parity of the number of fermions is conserved in the sector we consider, so $\text{exp}(i\pi\sum_{l=1}^N c_l^\dagger c_l)=1$. This quadratic fermion Hamiltonian can be diagonalized by the canonical transformation. We write Eq.(\ref{fh}) in matrix form as $\tilde{\mathscr{H}}=\frac{1}{2}C^\dagger M C$, where $C^\dagger=(c^\dagger_1,c^\dagger_2,\cdots,c^\dagger_N,c_1,c_2,\cdots,c_N)$, and
 \begin{eqnarray}
 M=\left(
     \begin{array}{cc}
       A & B \\
       -B & -A \\
     \end{array}
   \right)
 \end{eqnarray}
 where $A$ and $B$ are $N\times N$ matrix with elements given by $A_{i,i}=2J_i$, $A_{i,i+1}=-\lambda_i$, $A_{i+1,i}=-\lambda_i$, $B_{i,i+1}=\lambda_i$, $B_{i+1,i}=-\lambda_i$. It is worth noting that the sign of boundary terms needs to be changed caused by the even parity of particle numbers. The matrix $M$ can be diagonalized numerically after the orthogonal transformation
 \begin{eqnarray}\label{diagonalization}
 \left(
   \begin{array}{cc}
     A & B \\
     -B & -A \\
   \end{array}
 \right)
 =\left(
    \begin{array}{cc}
      g^T & h^T \\
      h^T & g^T \\
    \end{array}
  \right)
 \left(
   \begin{array}{cc}
     \omega & 0 \\
     0 & -\omega \\
   \end{array}
 \right)
 \left(
   \begin{array}{cc}
     g & h \\
     h & g \\
   \end{array}
 \right).
 \end{eqnarray}
 The spectrum of diagonal matrix $\omega$ consists of the elementary excitations of $\tilde{\mathscr{H}}$.

 To calculate the many-spin correlation function in Eq.(\ref{purity}), we can define `majorana like' operators $a_j=c_j^\dagger+c_j$ and $b_j=c_j^\dagger-c_j$ with $a^2_j=1$ and $b^2_j=-1$. Then we represent the spin operators as $\tau^z_j=a_jb_j$ and $\tau^x_j\tau^x_{j+1}=b_ja_{j+1}$. It is revealed that the modula square of the correlation function has the form of $|\langle\tilde{\psi}(t)|\cdots a_l\cdots a_s\cdots b_u \cdots b_v \cdots|\tilde{\psi}(t)\rangle|^2$, or $|\langle\tilde{\psi}(0)|\cdots a^H_l(t)\cdots a^H_s(t)\cdots b^H_u(t)\cdots b^H_v(t) \cdots|\tilde{\psi}(0)\rangle|^2$ written in Heisenberg picture. Owing to Wick's theorem, this many-fermion correlation function can be expressed as a Pfaffian \cite{barouch:1971}. So All we need to know, in the end, are three types of two-point correlation function:
\begin{eqnarray}\label{Geq}
G_{ij}(t)&=&\langle\tilde{\psi}(0)|a^H_i(t)b^H_j(t)|\tilde{\psi}(0)\rangle,\nonumber\\
G^A_{ij}(t)&=&\langle\tilde{\psi}(0)|a^H_i(t)a^H_j(t)|\tilde{\psi}(0)\rangle,\nonumber\\
G^B_{ij}(t)&=&\langle\tilde{\psi}(0)|b^H_i(t)b^H_j(t)|\tilde{\psi}(0)\rangle.
\end{eqnarray}

As mentioned before, $|\tilde{\psi}(0)\rangle$ is the vacuum state, namely $c_j|\tilde{\psi}(0)\rangle=0$ for any $j$. So we can expand $a^H_i(t)$, $b^H_i(t)$ by
\begin{eqnarray}
a^H_i(t)=\sum_j\phi^\ast_{ij}(t)c^\dagger_j+\phi_{ij}(t)c_j,\nonumber\\
b^H_i(t)=\sum_j\psi^\ast_{ij}(t)c^\dagger_j-\psi_{ij}(t)c_j.
\end{eqnarray}
Substitute the above equations in to Eq.(\ref{Geq}), we get
\begin{eqnarray}\label{Geq2}
G_{ij}(t)&=&\sum_l\phi_{il}(t)\psi_{jl}^\ast(t)=\left[\phi(t)\psi^\dagger(t)\right]_{ij},\nonumber\\
G^A_{ij}(t)&=&\sum_l\phi_{il}(t)\phi_{jl}^\ast(t)=\left[\phi(t)\phi^\dagger(t)\right]_{ij},\nonumber\\
G^B_{ij}(t)&=&-\sum_l\psi_{il}(t)\psi_{jl}^\ast(t)=-\left[\psi(t)\psi^\dagger(t)\right]_{ij}.
\end{eqnarray}

The matrixes $\phi(t)$ and $\psi(t)$ can be evaluated by solving the Heisenberg equation $i\frac{d}{dt}c^H_i(t)=\left[c^H_i(t),\tilde{\mathscr{H}}\right]=\sum_j A_{ij}c^H_{j}(t)+B_{ij}c^{H}_{j}(t)^\dagger $. Together with the expansion $c^H_i(t)=\sum_jg_{ij}(t)c_j+h^{\ast}_{ij}(t)c^{\dagger}_j$, we get the matrix equation
\begin{eqnarray}
\!\!\!\!\!i\frac{d}{dt}\left(\!
               \begin{array}{cc}
                 g(t) & h^{\ast}(t) \\
                 h(t) & g^{\ast}(t) \\
               \end{array}
             \!\!\right)=
             \left(\!\!
               \begin{array}{cc}
                 A & B \\
                 -B & -A \\
               \end{array}
             \!\!\right)
             \left(\!
               \begin{array}{cc}
                 g(t) & h^{\ast}(t) \\
                 h(t) & g^{\ast}(t) \\
               \end{array}
             \!\!\right).
\end{eqnarray}
We can solve the above equation by substituting Eq.(\ref{diagonalization}) into it and combining with the initial condition $g(t)=I$ and $h(t)=0$. The solution is
\begin{eqnarray}
\!\left(\!\!
  \begin{array}{cc}
    g(t) & \!h^{\ast}(t) \\
    h(t) & \!g^{\ast}(t) \\
  \end{array}
\!\!\!\right)\!=\!
\left(
  \begin{array}{cc}
    g^T & h^T \\
    h^T & g^T \\
  \end{array}
\right)
\left(
  \begin{array}{cc}
    e^{-i\omega t} & 0 \\
    0 & e^{i\omega t} \\
  \end{array}
\right)
\left(
  \begin{array}{cc}
    g & h \\
    h & g \\
  \end{array}
\right).\nonumber\\
\ignore{=
\left(\!
  \begin{array}{cc}
    g^T\!e^{-i\omega t}\!g\!+\!h^T\!e^{i\omega t}\!h & g^T\!e^{-i\omega t}\!h\!+\!h^T\!e^{i\omega t}\!g\\
    h^T\!e^{-i\omega t}\!g\!+\!g^T\!e^{i\omega t}\!h & h^T\!e^{-i\omega t}\!h\!+\!g^T\!e^{i\omega t}\!g\\
  \end{array}
\!\!\right).\nonumber\\}
\end{eqnarray}
Applying $\phi(t)=g(t)+h(t)$ and $\psi(t)=g(t)-h(t)$, we have
\begin{eqnarray}\label{phipsi}
\phi(t)&=&\phi^T\text{cos}\omega t\phi-i\phi^T\text{sin}\omega t\psi\nonumber\\
\psi(t)&=&\psi^T\text{cos}\omega t\psi-i\psi^T\text{sin}\omega t\phi,
\end{eqnarray}
where $\phi=g+h$ and $\psi=g-h$. Combining with Eq.(\ref{phipsi}) and Eq.(\ref{Geq2}) we can compute the pfaffian of each may-spin correlation function by reducing it to a determinant.


\begin{thebibliography}{64}
\bibitem{note1} {We note here that these constraints dose not change the general result. We illustrate in \cite{Yu2016} that the density matrix has the same form in different topological sectors, thus the entanglement entropy and topological entropy are unchanged.}
\end{thebibliography}
\end{document}